\documentclass[12pt, amsfonts]{article}
\begin{document}
\def\p {{\partial}}
\def\n {{\nu}}
\def\m {{\mu}}
\def\a {{\alpha}}
\def\bt {{\beta}}
\def\f {{\phi}}
\def\th {{\theta}}
\def\g {{\gamma}}
\def\eps {{\epsilon}}
\def\e {{\psi}}
\def\k {{\chi}}
\def\la {{\lambda}}
\def\na {{\nabla}}
\def\bn {\begin{eqnarray}}
\def\en {\end{eqnarray}}
\title{Path integral quantization of Yang- Mills theory\footnote{e-mail:
$sami_{-}muslih$@hotmail.com}} \maketitle
\begin{center}
\author{S. I. MUSLIH\\\it{Department of Physics Al-Azhar University
Gaza, Palestine}}
\end{center}
\hskip 5 cm

$\mathbf{Summary}$.- Path integral formulation based on the
canonical method is discussed. Path integral for Yang-Mills
theory is obtained by this procedure. It is shown that gauge
fixing which is essential procedure to quantize singular systems
by Faddeev's and Popov's method is not necessary if the canonical
path integral formulation is used.

PACS 0.3.50. Kk- Other special classical field theories.

PACS 0.3.70- Theory of quantized field.

PACS 11.15- Gauge field theories.

PACS 11.10. Ef- Lagrangian and Hamiltonian approach.

\newpage

\section{Introduction}

Quantization of classical systems can be achieved by the
canonical quantization method [1]. If we ignore the ordering
problems, it consists in replacing the classical Poisson brackets,
by quantum commutators when classically all the states on the
phase space are accessible. This is no longer correct in the
presence of constraints. An approach due to Dirac [2] is widely
used for quantizing the constrained Hamiltonian systems [3,4].
Physicists start to use the canonical method because of some
important properties of quantum theory, such as unitarity and
positive definiteness of the metric, may be deduced easily.

The alternative quantization scheme for constrained systems is
the path integral quantization. It is important because it serves
as a basis to develop perturbation theory and to find out the
Fynman rules. The path integral quantization of singular theories
with first class constrains in canonical gauge was given by
Faddeev and Popov [5,6]. The generalization of the method to
theories with second class constraints is given by Senjanovic
[7]. Moreover, Fradkin and Vilkovisky [8,9] considered
quantization to bosonic theories with first class constraints and
it is extension to include fermions in the canonical gauge. More,
Gitman and Tyutin [4] discussed the canonical quantization of
singular systems as well as the Hamiltonian formalism of gauge
theories in an arbitrary gauge.

When the dynamical system possesses some second class constraints
there exists another method given by Batalain and Fradkin [10]:
the BFV-BRST operator quantization method. One enlarges the phase
space such that the original second class constraints became
converted into the first class ones, so that the number of
degrees of freedom remain unaltered.

Recently another method based on the canonical method [11-14] is
introduced to obtain the path integral quantization of singular
systems [15-16]. The starting point of this method is the
variational principle. The Hamiltonian treatment of the
constrained systems leads to obtain the equations of motion as
total differential equations in many variables which require the
investigation of integrability conditions. The equations of
motion are integrable if the corresponding system of partial
differential equations is a Jacobi system [11]. In this case one
can construct a valid and a canonical phase space coordinates .
The path integral then obtained as an integration over the the
canonical phase space coordinates with the action is obtained
directly from the equations of motion. Some applications of the
canonical path integral method are given in references [17-19]
and it shown that gauge fixing is not necessary to quantize
singular systems if the canonical method is used, no need to
enlarge the phase space, no need to introduce delta functions, as
well as no ambiguous determinant will appear. This new approach,
due to its very recent development, has been applied to very few
examples [17-19] and a better understanding of its features, its
advantages in the study of singular systems when compared to
other conventional methods [7-9] is still necessary.

This paper is arranged as follows: A brief information on the
canonical path integral method is given section $\mathbf{2}$.
Path integral of a pure Yang-Mills theory is worked out in
section $\mathbf{3}$.
\section{A summary of the canonical path integral formulation}

As was discussed in previous papers [15,16], gauge fixing is not
necessary to quantize singular systems if the canonical path
integral formulation is used. Thus, it will be instructive to
give a brief discussion of this method.

The canonical method gives the set of Hamilton - Jacobi partial
differential equations [HJPDE] as

\bn
&&H^{'}_{\a}(t_{\bt}, q_a, \frac{\p S}{\p q_a},\frac{\p S}{\p
t_a}) =0,\nonumber\\&&\a, \bt=0,n-r+1,...,n, a=1,...,n-r,\en where
\begin{equation}
H^{'}_{\a}=H_{\a}(t_{\bt}, q_a, p_a) + p_{\a},
\end{equation}
and $H_{0}$ is defined as
\bn
 &&H_{0}= p_{a}w_{a}+ p_{\m} \dot{q_{\m}}|_{p_{\n}=-H_{\n}}-
L(t, q_i, \dot{q_{\n}},
\dot{q_{a}}=w_a),\nonumber\\&&\m,~\n=n-r+1,...,n. \en

The equations of motion are obtained as total differential
equations in many variables as follows:

\bn
 &&dq_a=\frac{\p H^{'}_{\a}}{\p p_a}dt_{\a},\;
 dp_a= -\frac{\p H^{'}_{\a}}{\p q_a}dt_{\a},\;
dp_{\bt}= -\frac{\p H^{'}_{\a}}{\p t_{\bt}}dt_{\a}.\\
&& dz=(-H_{\a}+ p_a \frac{\p
H^{'}_{\a}}{\p p_a})dt_{\a};\\
&&\a, \bt=0,n-r+1,...,n, a=1,...,n-r\nonumber \en where
$z=S(t_{\a};q_a)$. The set of equations (4,5) is integrable [11]
if

\bn &&dH^{'}_{0}=0,\\
&&dH^{'}_{\m}=0,\;\; \m=n-p+1,...,n. \en If condition (6,7) are
not satisfied identically, one considers them as new constraints
and again testes the consistency conditions. Hence, the canonical
formulation leads to obtain the set of canonical phase space
coordinates $q_a$ and $p_a$ as functions of $t_{\a}$, besides the
canonical action integral is obtained in terms of the canonical
coordinates.The Hamiltonians $H^{'}_{\a}$ are considered as the
infinitesimal generators of canonical transformations given by
parameters$t_{\a}$ respectively. In this case, the path integral
representation may be written as [15,16]

\bn &&\langle Out|S|In\rangle=\int \prod_{a=1}^{n-p}dq^{a}~dp^{a}
[\exp i \{\int_{t_{\a}}^{{t'}_{\a}}(-H_{\a}+ p_a\frac{\p
H^{'}_{\a}}{\p p_a})dt_{\a}\}],\nonumber\\&&a=1,...,n-p,
\;\;\;\a=0,n-p+1,...,n. \en

One should notice that the path integral (8) is an integration
over the canonical phase-space coordinates $q_{a}$ and $p_{a}$.

\section{ Quantization of Yang-Mills theory}

In this section we shall consider the path integral quantization
of Yang-Mills theory and demonstrate the fact that gauge fixing
problem is solved if the canonical path integral method is used.
The action of this theory is given as
\begin{equation}
S[A_{\m}]=\int {\cal L}~d^{4}x,
\end{equation}
where the Lagrangian density is
\begin{equation}
{\cal L}= -\frac{1}{4} F_{\m\n}^{a} F_{a}^{\m\n},\;\;\m,\;\n= 0,
1, 2, 3,
\end{equation}
with the field strength $F_{\m\n}^{a}$ defined as
\begin{equation}
F_{\m\n}^{a}= \p_{\m} A_{\n}^{a} - \p_{\n} A_{\m}^{a} + g f^{abc}
A_{\m}^{b}A_{\n}^{c},
\end{equation}
where $f^{abc}$ are the structure constants of the Lie-Algebra
and $g$ represents the coupling constant. This action is
invariant under gauge transformations.

The momenta conjugated to the fields $A_{\m}^{a}$ are defined as
\begin{equation}
\pi_{a}^{\m}= \frac{\p {\cal L}}{\p(\p_{0} A_{\m}^{a})} =
F_{a}^{\m 0}.
\end{equation}
The non vanishing Poisson brackets are
\begin{equation}
\{A_{\m}^{a}(x),
\pi_{b}^{\n}(x')\}=\delta_{\m}^{\n}\delta_{b}^{a}\delta^{(3)}(x'-x).
\end{equation}
Upon quantization, these brackets have to be converted into
proper commutators.

The spatial components read as
\begin{equation}
\pi_{a}^{i}= F_{a}^{i0},\;\;\;i=1, 2, 3,
\end{equation}
and the time component
\begin{equation}
\pi_{a}^{0}= 0,
\end{equation}
is the primary constraint. Defining the covariant derivatives
$D_{i}$ as
\begin{equation}
D_{i}x^{a}= {\p}_{i}x^{a} + g f^{abc}A_{i}^{b}x^{c},
\end{equation}
in this case Eqn. (14) leads us to express the velocities ${\dot
{A_{i}^{a}}}$ in terms of the momenta $\pi_{a}^{i}$ as
\begin{equation}
{\dot {A_{a}^{i}}} = -({\pi}_{a}^{i} - D^{i}A_{a}^{0}).
\end{equation}

The total canonical Hamiltonian takes the form
\begin{equation}
H_{0}= \int[\frac{1}{4} F_{ij}^{a}
F_{a}^{ij}-\frac{1}{2}\pi_{a}^{i}\pi_{i}^{a}-
D_{i}\pi_{a}^{i}A_{0}^{a} + \p_{i}(\pi_{a}^{i}A_{0}^{a})]d^{3}x.
\end{equation}

Starting from the Hamiltonian defined in (18) and making use of
(2), (15), the set of Hamilton -Jacobi partial differential
equations reads as \bn&&{H'}_{0}= \pi_{4} +
H_{0},\;\;\;\;\;\pi_{4}=\frac{\p S}{\p t},\\
&&{H'}_{a}= \pi_{a}^{0}= 0,\;\;\;\;\;\;\;\pi_{0}^{a}=\frac{\p
S}{\p A_{a}^{0}}. \en The total differential equations can be
written as \bn&& dA_{a}^{i}= -({\pi}_{a}^{i} -
D^{i}A_{a}^{0})dt,\\
&&d{\pi}_{a}^{i}= (D_{j} F_{a}^{ji} +
gf_{abc}\pi_{b}^{i}A_{c}^{0})dt,\;\;\;j=1, 2, 3,\\
&&d\pi_{a}^{0}= (D_{i}\pi_{a}^{i})dt,\\
&&d\pi^{4}=0.
\en

To check whether the set of equations (21-24) is integrable or
not, let us consider the variation of (19) and (20). In fact
\begin{equation}
d{H'}_{0}= F_{1}^{a} dA_{0}^{a},
\end{equation}
where
\begin{equation}
F_{1}^{a}= - D_{i}\pi_{a}^{i}.
\end{equation}
Since $ F_{1}^{a}$ is not identically zero, we consider it as a
new constraint, and one should consider the variation of
$F_{1}^{a}$ too. Calculation shows that it lead to the constraint
\begin{equation}
g f_{abc}A_{0}^{b}D_{i}\pi_{c}^{i}=0.
\end{equation}
The variation of ${H'}_{a}$ is zero simply because it is equal to
$-F_{1}^{a}$.

The set of equations (21-24) is integrable, Hence, the canonical
phase-space coordinates $A_{a}^{i}$ and $\pi_{a}^{i}$ are
obtained in terms of parameters $t$ and $A_{0}^{a}$. Making use
of equation (5), the canonical action integral is calculated as
\begin{equation}
z= \int[-\frac{1}{4} F_{ij}^{a}
F_{a}^{ij}+\frac{1}{2}\pi_{a}^{i}\pi_{i}^{a}+
D_{i}\pi_{a}^{i}A_{0}^{a} - \p_{i}(\pi_{a}^{i}A_{0}^{a}) +
\pi_{a}^{i}\p_{0}A_{i}^{a} ]d^{4}x.
\end{equation}
Making use of (28) and (8) we obtain the path integral as \bn
\langle Out|S|In\rangle&&=
\int\prod_{i,a}DA_{i}^{a}~D\pi_{i}^{a}\exp i\{ \int[-\frac{1}{4}
F_{ij}^{a}
F_{a}^{ij}+\frac{1}{2}\pi_{a}^{i}\pi_{i}^{a}+\nonumber\\&&
D_{i}\pi_{a}^{i}A_{0}^{a} - \p_{i}(\pi_{a}^{i}A_{0}^{a}) +
\pi_{a}^{i}\p_{0}A_{i}^{a} ]d^{4}x\}. \en

One should notice that the path integral (28) has no singular
nature. In fact the path integral expression (28) is an
integration over the canonical phase space coordinates
$A_{i}^{a}$ and $\pi_{i}$.

We know that for a system with $n$ degrees of freedom and with $
r$ first class constraints $\f^{\a}$ the path integral
representation is given by Faddeev [5,6] as
\begin{equation}
\langle Out|S|In\rangle=\int \prod_{t} d{\m}(q_j, p_j)\exp i
\{\int_{-\infty}^{\infty}dt(p_j \dot{q_j} - H_0)\},\;\;j=1,...,n,
\end{equation}
where the measure of integration is given as
\begin{equation}
d {\m}= det|\{\f^{\a},
\k^{\bt}\}|\prod_{\a=1}^{r}\delta(\k^{\a})\delta(\f^{\a})\prod_{j=1}^{n}dq^{j}
dp_{j},
\end{equation}
and $\k^{\a}$ are $r$- gauge constraints.

If we perform now the usual path integral quantization [5,6]
using (30) for system (9), one must choose two gauge fixing
conditions to obtain the path integral quantization over the
canonical phase space coordinates $A_{i}^{a}$ and $\pi_{i}^{a}$.

\section{ Conclusion}

Path integral quantization of Yang-Mills theory is obtained using
the canonical path integral formulation [15,16]. In this approach,
since the integrability conditions $d{H'}_{0}=0, d{H'}_{a}=0$ are
satisfied, this system is integrable. Hence, the canonical phase
space coordinates $A_{i}^{a}, \pi_{i}^{a}$ are obtained in terms
of parameters $(t, A_{0}^{a})$. In this case the path integral,
then follows directly as given in (29) without using any gauge
fixing conditions. In the usual formulation [5-10], one has to fix
a gauge to obtain the path integral over the canonical variables.

As a conclusion, it is obvious that one does not fix any gauge if
the canonical path integral formulation is used. If the system is
integrable, one can construct canonical phase space coordinates
$q_{a}$ and $p_{a}$ in terms of parameters $t_{\a}$. Besides the
canonical action integral is naturally obtained from the
equations of motion without introducing Lagrange multipliers. In
this case the path integral quantization is obtained directly as
an integration over the canonical phase space coordinates.

Another point to be specified is that as we mentioned in the
introduction, the canonical path integral approach is a new
method and we still lack a complete analysis of relation between
the procedure in this method for constrained systems and the
traditional ones, especially with Faddeev's and Senjanov's
methods.

\end{document}